\documentclass[amsmath,amssymb,pra,twocolumn,groupaddress]{revtex4-2}
\usepackage{graphicx}
\usepackage{dcolumn}
\usepackage{bm}
\usepackage[colorlinks,linkcolor=blue,anchorcolor=blue,citecolor=blue,]{hyperref}
\usepackage{xcolor}

\begin{document}

\title{Lamb dip of a Quadrupole Transition in H$_2$}

\author{F. M. J. Cozijn}
\affiliation{Department of Physics and Astronomy, LaserLab, Vrije Universiteit\\
De Boelelaan 1081, 1081 HV Amsterdam, The Netherlands}

\author{M. L. Diouf}
\affiliation{Department of Physics and Astronomy, LaserLab, Vrije Universiteit\\
De Boelelaan 1081, 1081 HV Amsterdam, The Netherlands}

\author{W. Ubachs}
\affiliation{Department of Physics and Astronomy, LaserLab, Vrije Universiteit\\
De Boelelaan 1081, 1081 HV Amsterdam, The Netherlands}
 \email{w.m.g.ubachs@vu.nl}

\date{\today}

\begin{abstract}
\noindent
The saturated absorption spectrum of the hyperfine-less S(0) quadrupole line in the (2-0) band of H$_2$ is measured at $\lambda=1189$ nm, using the NICE-OHMS technique under cryogenic conditions (72~K). 
It is for the first time that a Lamb dip of a molecular quadrupole transition is recorded.
At low (150-200 W) saturation powers a single narrow Lamb dip is observed, ruling out an underlying recoil doublet of 140 kHz. 
Studies of Doppler-detuned resonances show that the red-shifted recoil component can be made visible for low pressures and powers, and prove that  the narrow Lamb dip must be interpreted as the blue recoil component. 
A transition frequency of 252\,016\,361\,164\,(8) kHz is extracted, which is off by -2.6 (1.6) MHz from molecular quantum electrodynamical calculations therewith providing a challenge to theory.

\end{abstract}

\maketitle

The hydrogen molecule has been a test ground for the development of molecular quantum mechanics for almost a century~\cite{Heitler1927}. 
In the recent decade the level of precision has been accelerated in benchmark experimental studies focusing on its dissociation and ionization energies~\cite{Cheng2018,Holsch2019}, now reaching perfect agreement with first principles calculations based on four-particle variational calculations and including relativistic and quantum electrodynamic (QED) effects~\cite{Puchalski2019b,Puchalski2019}.
The target of activity has in part shifted to measurements of the vibrational  quantum in the hydrogen molecule.
In the HD isotopologue vibrational transitions were detected in saturation.
Lamb dip spectroscopy could be performed at high precision due to the weak dipole moment in this heteronuclear species~\cite{Tao2018,Cozijn2018}.
However, in these studies a problem of extracting rovibrational transition frequencies surfaced. 
Observed asymmetric lineshapes were interpreted in various ways, in terms of underlying hyperfine structure and cross-over resonances~\cite{Diouf2019}, of Fano-type interferences~\cite{Hua2020}, 
and of effects of standing waves in the optical cavity~\cite{Lv2022}.
This situation, imposing unclarity on the extraction of energy separations between quantum levels, has halted further progress in the precision metrology of HD, although a focused activity remains~\cite{Cozijn2022,Liu2022}.

In the homonuclear H$_2$ species selection rules govern that only quadrupole transitions are allowed and those are two orders of magnitude weaker than the dipole absorption transitions in HD~\cite{Kassi2011}.
Vibrational transitions in H$_2$ have been probed in Doppler-broadened spectroscopy~\cite{Bragg1982,Hu2012,Kassi2014}, through combination differences of Doppler-free electronic transitions~\cite{Dickenson2013}, and recently via stimulated Raman scattering~\cite{Lamperti2022}.
While all rovibrational levels in HD are subject to complex hyperfine structure induced by the magnetic moment of both H and D nuclei, H$_2$ has the advantage that the levels in para-H$_2$ exhibit no hyperfine substructure. 
Until today no saturation spectroscopy has been performed on quadrupole transitions, neither in H$_2$ nor in any other molecule.

For performing saturation spectroscopy of an extremely weak quadrupole transition a novel setup was built as an upgrade from the setup used for the HD experiments~\cite{Cozijn2018,Diouf2019}.
The optical cavity is redesigned to suppress vibrations and attached to a cryo-cooler to reach temperatures in the range 50-300 K. At the typical operation range of 72 K a larger fraction of the population is condensed into the H$_2$ $J=0$ ground level, while the transit-time through the intracavity laser beam is reduced.
The laser, an external-cavity diode laser with a tapered amplifier running at 1189 nm, is locked to the cavity for short-term stability and to a frequency-comb-laser for sub-kHz accuracy in long-term measurements, thus also providing the absolute frequency scale. 
This stability allows for long time averaging over multiple scans.
The 371~mm hemispherical resonator is equipped with highly reflective (R $>$ 0.99999) mirrors of which the concave mirror has an ROC of 2~m. This yields a finesse of 350,000, an intracavity circulating power of up to 10~kW, and a beam waist of 542 $\mu$m.
Further details on the cryogenic NICE-OHMS spectrometer will be given in a forthcoming publication~\cite{Cozijn-CC}.

\begin{figure}[t]
\begin{center}
\includegraphics[width=0.98\linewidth,height=0.22\textheight]{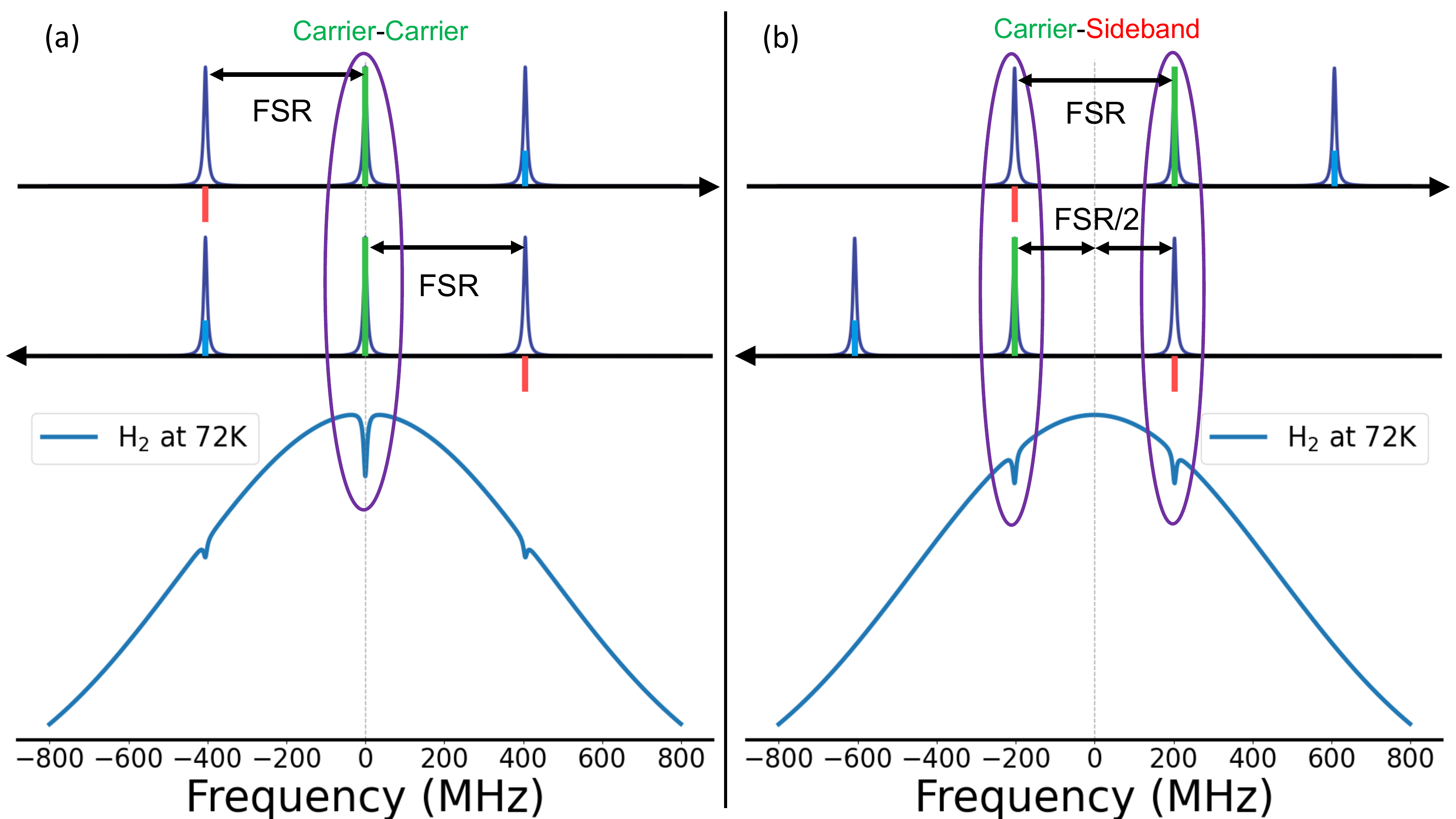}
\caption{\label{NO_signals}
Possible interactions of the three fields (carrier and two sidebands) inside the cavity with a molecular resonance. Panel (a) represents the generic Lamb dip generated by the counterpropagating carrier beam on resonance. Additional holes are burned in the Doppler broadened profile with the combined interaction of the red/blue sidebands. In panel (b), the carrier is detuned off resonance by $f_m/2$ (or FSR/2). In that condition, one of the sidebands (in this example red) interacts with the counterpropagating carrier and consequently burns holes at $\pm$ FSR/2. }
\end{center}
\end{figure}

Detection is based on the technique of noise-immune cavity-enhanced optical heterodyne molecular spectroscopy (NICE-OHMS)~\cite{Ma1999,Foltynowicz2008a,Axner2014a} using sideband modulation of the carrier wave, at frequency $f_c \pm f_m$ with $f_m=404$ MHz, matching the free-spectral-range (FSR) of the cavity for generating the heterodyne NICE-OHMS signal. The carrier and the two generated sidebands are locked to the cavity with Pound-Drever-Hall and DeVoe-Brewer~\cite{Devoe1984} stabilization, respectively. 
Consequently, the three beams counterpropagate inside the cavity and interact with the molecules present, giving rise to various sub-Doppler spectroscopic signals from which two possible schemes are shown in Fig.~\ref{NO_signals}. 
In panel (a), where the carrier is set on top of the resonance center, the counterpropagating carrier beams burn a hole in the center of the velocity distribution (at $v_{z}=0$) and generate the generic Lamb dip signal. 
Additionally, saturation conditions are formed by the red/blue sidebands interacting simultaneously on molecules with velocities $k\cdot v_{z}=\pm f_m$ and burning holes at their respective positions \cite{22DiToScCo}. However, this effect is typically negligible for the weakly saturating regime and for conditions of low sideband power. 

In panel (b), on the other hand, the laser is detuned from the molecular resonance by $f_m/2=202$~MHz (or FSR/2). Here, one of the sidebands (the red sideband in this example) in combination with the counterpropagating carrier beam interact with molecules with velocities $k\cdot v_{z}=\pm f_m/2$. 
As for this velocity class both beams are in resonance, a pump-probe scheme is formed, resulting in Doppler-detuned saturation signals. 
Since the required detuning is exactly known and the resulting Doppler-shift is equal, it can be seen as an alternative scheme for Doppler-free spectroscopy as only the known detuning needs to be considered to extract the transition frequency. The novelty of this scheme is that the ordinary on-resonance strong standing wave, present for the usual carrier-carrier saturation, is now converted to mostly a travelling wave due to the low intensity sideband. 
This allows to mitigate possible effects of the strong on-resonance standing wave.

In addition to the sideband modulation for the heterodyne signal, slow wavelength modulation of the cavity length is applied at 395 Hz with a peak-to-peak amplitude of 50 kHz. This allows for lock-in detection, where demodulation at the first derivative $(1f)$ is applied.  The $1f$  profile function is defined as a derivative of a typical dispersive Lorentzian profile~\cite{Foltynowicz2009a}
\begin{equation}
    f(\nu)_{1f} = \frac{4\,A\left[\Gamma^2 - 4(\nu - \nu_0)^2 \right]}{\left[ \Gamma^2 + 4 (\nu - \nu_0)^2 \right]^2},
    \label{Eq-NO_1f}
\end{equation}
where the adjustable parameters are the line position $\nu_0$, the line intensity $A$, and the width  $\Gamma$.

\begin{figure}[b]
\begin{center}
\includegraphics[width=1.1\linewidth,height=0.2\textheight]{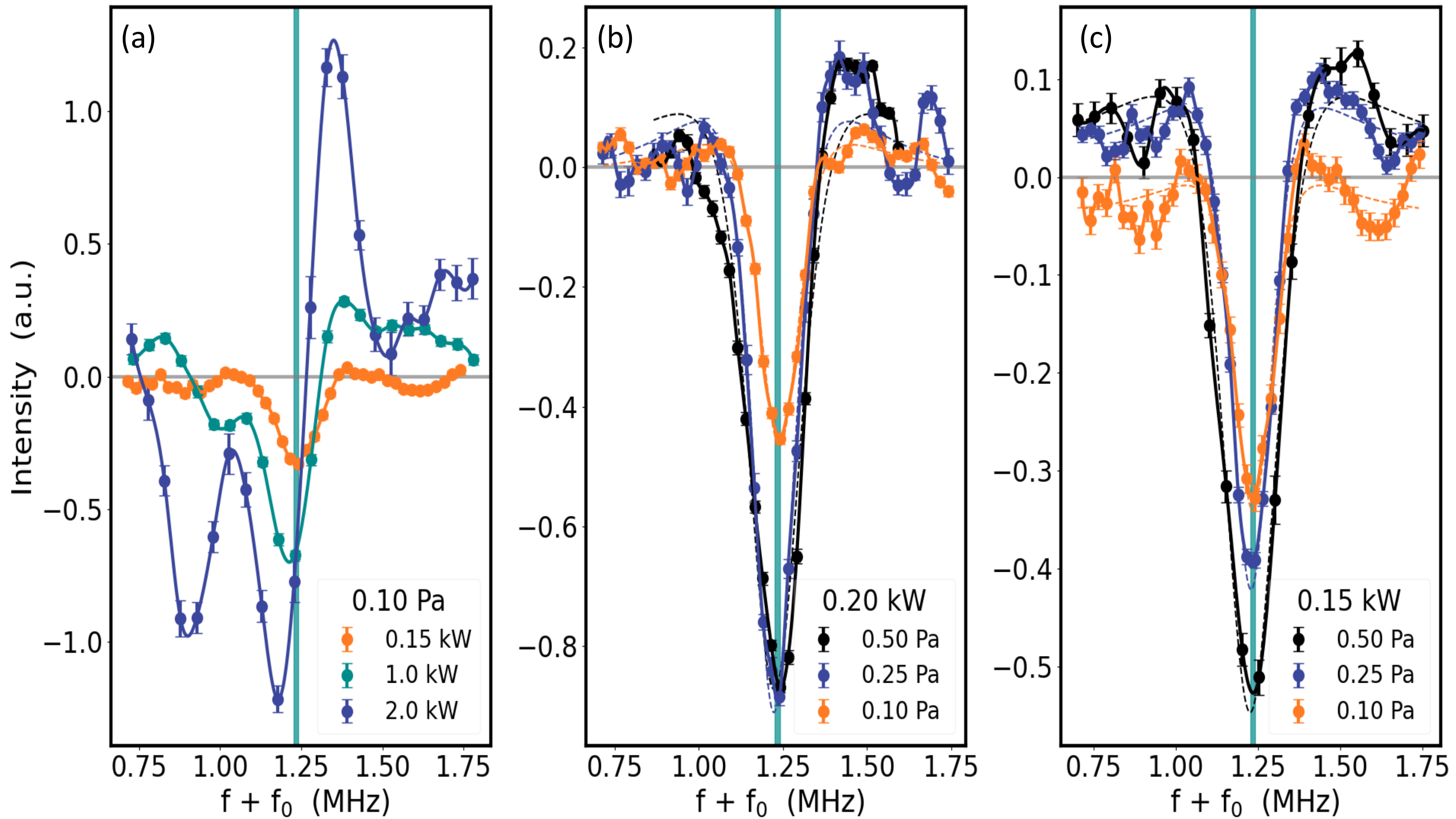}
\caption{\label{H2_S-spectra}
Recorded spectra of the measured Lamb dip for the S(0) (2-0) quadrupole transition in H$_2$ at 72 K at (a) 0.10 Pa for different circulating power as indicated. Panel (b) and (c) respectively show the measured spectra at the lower circulating powers of 150 Watt and 200 Watt for different pressure indicated. A  $1f$ dispersive Lorentzian (Eq.~\ref{Eq-NO_1f}) was superimposed on the measured spectra. The absolute frequency scale is given via $f_0 = 252\,016\,360$ MHz .
}
\end{center}
\end{figure}

Measurements of the S(0) (2-0) line were performed in the cryo-NICE-OHMS setup under a variety of conditions of intracavity power and pressure.
Since the quadrupole transition is extremely weak, with a line strength $S=1.6 \times 10^{-27}$ cm/molecule, and Einstein coefficient $A=1.3 \times 10^{-7}$ s$^{-1}$~\cite{Kassi2014}, it was anticipated that extreme powers would be required to obtain a saturation signal. 
At high power the spectra recorded displayed complex lineshapes, as shown in panel (a) of  Fig.~\ref{H2_S-spectra}, reminiscent of the dispersive line profiles observed in HD ~\cite{Diouf2019,Diouf2020,Hua2020,Lv2022}. 
Surprisingly, by lowering the power, the complex lineshapes at 2.0 kW turn to an asymmetric dispersive-like profile at 1.0 kW, to finally a symmetric profile at 150 W. 
Panel (b) and (c) show the symmetric Lamb dip obtained at the lower powers of 150 W and 200 W, where each individual measurement was obtained after 12 hours of averaging.  
A $1f$ dispersive Lorentzian (Eq.~\ref{Eq-NO_1f}) fit was then used on the symmetric profiles so as to extract relevant parameters such as the Lamb dip position and the linewidth.

\begin{figure}[t]
\begin{center}
\includegraphics[width=\linewidth,height=0.25\textheight]{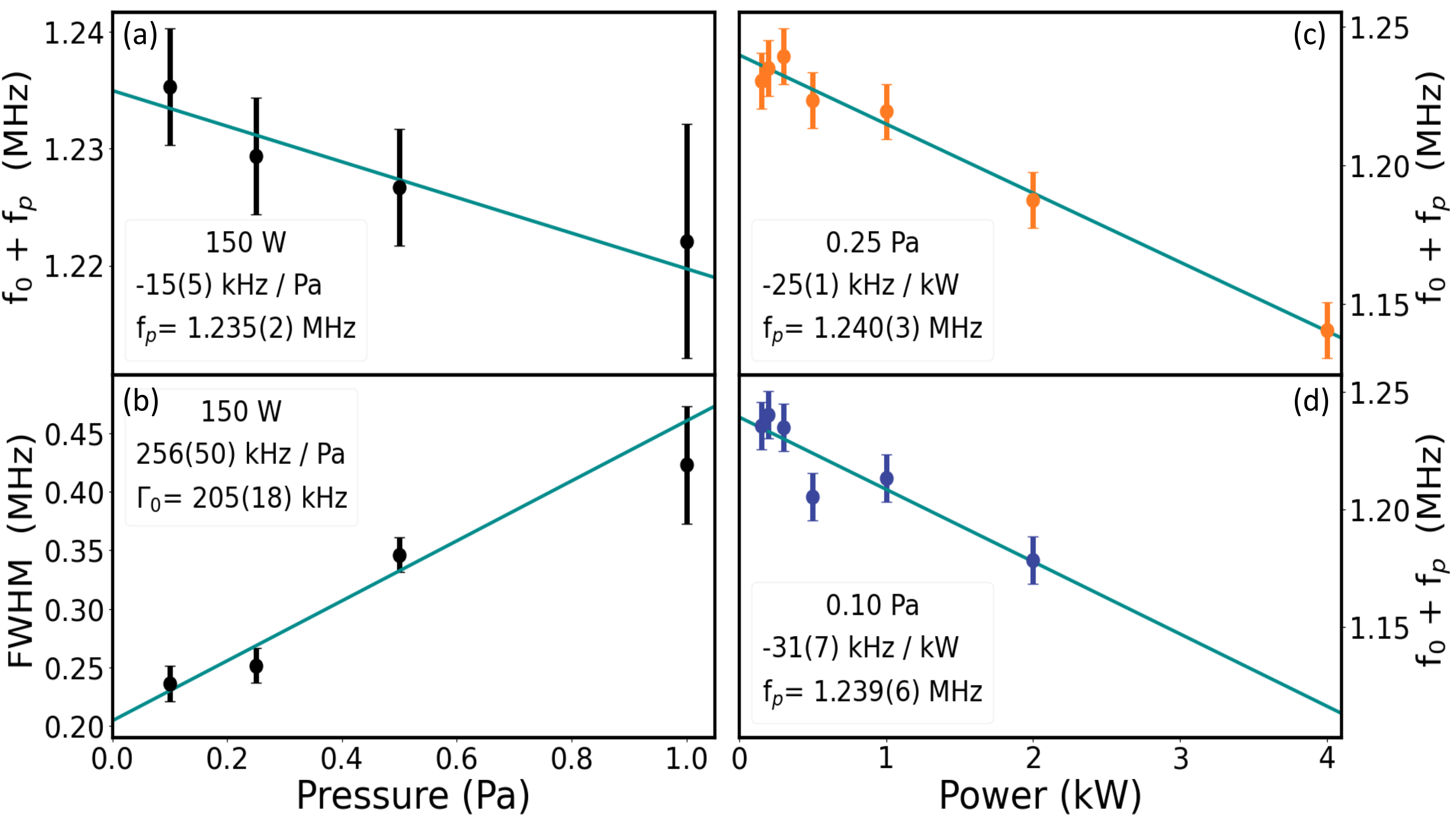}
\caption{\label{analysis}
Extracted positions of the S(0) Lamb dip at different powers and pressures and extrapolation to zero values. (a) Pressure dependence and shift at $P=150$ W; (b) Pressure dependent width at 150 W (FWHM); (c) and (d)  Power-dependent shifts at $p=0.25$ and $0.1$ Pa. The pressure- and power-dependent slopes are as indicated. The absolute frequency scale is given via $f_0 = 252\,016\,360$ MHz. 
}
\end{center}
\end{figure}

Large sets of data were obtained, mainly at 150 W, 200 W, and 300 W, where symmetric lines were observed, but also at higher powers.
The extracted positions of the Lamb dip were treated in a multivariate analysis yielding a transition frequency extrapolated to zero-pressure and zero-power of $f = 252\,016\,361\,234.4\,(7.3)$ kHz, which we will refer to as the 'generic Lamb dip' in the following. 
Some subsets of pressure-dependent (at 150 W) and power-dependent (at 0.25 Pa and 0.10 Pa) curves are shown in Fig.~\ref{analysis}.

Extrapolating the extracted widths to zero pressure yields a linewidth limit of 205 kHz (FWHM) for 150 W, which still overestimates the actual limit as dithering effects are not removed. 
These values are considerably smaller than the calculated 471 kHz transit-time width (for 72 K)~\cite{Shimoda} and can be attributed to the selection of cold molecules in the weakly saturating regime \cite{Bagayev1989}, as observed in our previous work on HD \cite{Cozijn2018}. The observed width corresponds to a most probable velocity of $v_{\rm mp}=335$ m/s and temperature of around 13 K. 

In order to accurately extract the transition frequency one needs to consider and correct for the known Doppler shifts and the resulting recoil from conservation of momentum. The total energy carried by a photon for making a transition or released from emission is expressed as~\cite{Demtroeder}
\begin{equation}
  E_{\rm photon} = h\nu_0 \pm \frac{h \vec{k}\cdot\vec{v}}{2\pi}  \pm \frac{(h\nu_0)^2}{2mc^2} - \frac{(h\nu_0)v^2}{2c^2}.
  \label{Eq-recoil}
\end{equation}
Here, $h\nu_0$ is the true energy difference between quantum levels. The second term is the first order Doppler shift and is equal to zero under conditions of saturation. The third term is the recoil shift, where the plus/minus sign refers to the case of absorption/stimulated emission.
The final term represents the second-order (relativistic) Doppler effect, which is as small as 160 Hz (for 13~K).

\begin{figure}[b]
\begin{center}
\includegraphics[width=\linewidth,height=0.25\textheight]{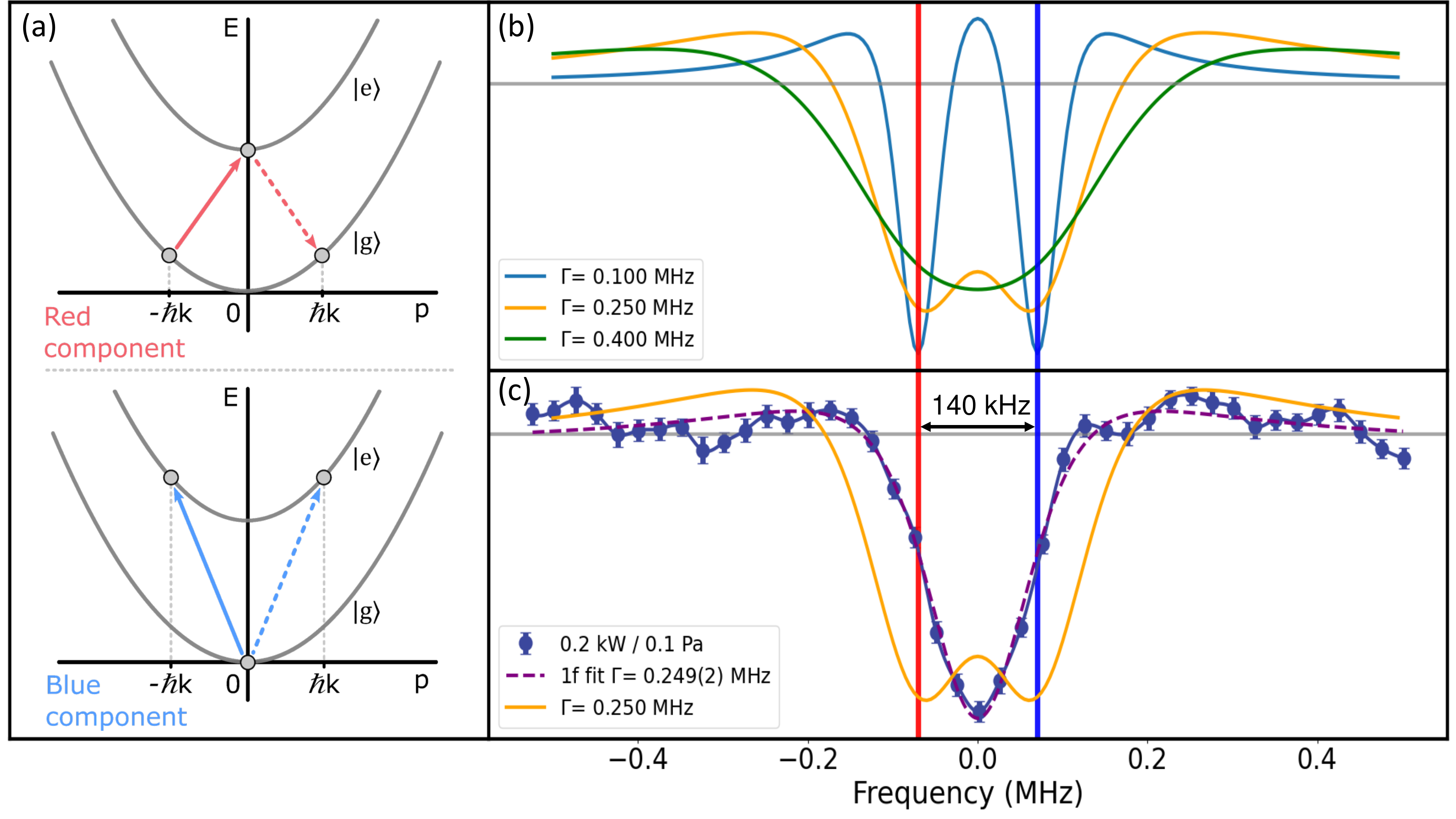}
\caption{\label{Recoil_model}
(a) Schematic of the blue (absorption) and red (stimulated emission) recoil components when the two counterpropagating lasers are close to a resonant transition $\vert e\rangle \leftarrow \vert g \rangle $. (b) Modeled profiles of a recoil doublet for the expected recoil splitting of 140 kHz and for varying levels of broadening. (c) A measurement of a 'generic Lamb dip' at a pressure 0.1 Pa, intracavity power of 0.2 kW and temperature 72 K. The observed symmetric lineshape cannot be modeled by any of the simulated profiles that include the 140 kHz recoil splitting while the simple $1f$ dispersive Lorentzian fit produces perfect agreement.
}
\end{center}
\end{figure}

In saturation spectroscopy two recoil components are associated with each quantum transition due to conservation of momentum \cite{Kolchenko1968}. A high-frequency (blue-detuned) component occurs for absorption from ground-state particles, and a low-frequency (red-detuned) component for stimulated emission from excited-state particles (Fig.~\ref{Recoil_model}). 
Both components will form individual Lamb dips as a characteristic recoil doublet, split at twice the recoil shift and centered around the resonance center. 
Despite the significance of recoil on extracting the transition frequency, it is often neglected in saturation spectroscopy as typically the observed linewidths are significantly larger than the recoil doublet splitting and thereby making the recoil doublet unresolvable. 
Nevertheless, there have been studies in which the recoil doublet has been successfully resolved in atoms~\cite{Barger1979,Riehle1988,Kurosu1992,Oates2005} and in molecules~\cite{Hall1976b,Alekseev1984,Bagayev1989,Bagayev1991}.

For the present case of H$_2$ and the transition frequency of the S(0) line this recoil amounts to 70 kHz and a total splitting of 140 kHz. 
As this is only marginally larger than the observed linewidth of 230 kHz, effects of the recoil splitting are expected to be visible on the observed lineshape. 
Model calculations are presented for the known recoil doublet splitting and a variety of widths that each component may obtain (Fig.~\ref{Recoil_model}). Comparison of simulated profiles with measurements reveal that the observed lineshape cannot be composed from both recoil components. In fact, in Fig.~\ref{Recoil_model}(b), a distinction is clearly seen as the measured Lamb dip does not display any hint of an unresolved recoil doublet and is perfectly fitted by a $1f$ Lorentzian profile composed of a single transition. This leads to the conclusion that the observed generic Lamb dip at low power cannot consist of both recoil components and that suppression of one of the recoil components has occurred.

There have been studies on suppression on both the red-shifted \cite{Riehle1988}  and blue-shifted \cite{Kurosu1992} recoil components. In either case this was performed by depopulating the upper-state or ground-state respectively through (optical) pumping. Also in an early observation of the resolved recoil doublet unequal intensity components were found, in which the red-shifted component was suppressed under some conditions \cite{Hall1976b}. 
In the theoretical derivation from Kol'chenko et al.~\cite{Kolchenko1968} it was found that the ratio of the depths of the recoil components are determined by the lifetimes of the states involved. 
From these observations and findings it can be reasoned that due to the two-step process of the stimulated emission scheme
and the typical lower lifetime of the excited state, the red-shifted recoil component is more easily suppressed.

In the case of H$_2$ the natural lifetimes of states are non-restrictive and other effects of relaxation must be considered.
Collisional effects, the finite transit times and effects of the strong standing wave can all be considered as effective methods of depopulation or dephasing. 
Compared to the previous studies where the recoil doublet was successfully resolved, our present study operates at around one to two orders of magnitude higher pressure. 
The most striking difference between this study and the previous studies on the CH$_4$ molecule is the use of extreme laser intensities to saturate the weak quadrupole transition. 
In our study, up to 10-12 orders of magnitude higher intensities~\cite{Bagayev1989,Bagayev1991} are present which can lead to significant standing wave effects in the optical resonator.

Effects of a strong standing wave on neutral molecules had been theoretically explored in the past by Letokhov and Chebotayev~\cite{LetokhovChebotayev}. The finite polarizability of molecules leads to an axial striction force due to the strong electric field gradient, imposing axial velocity modulation, or ultimately, even axial trapping of molecules. The resulting velocity modulation can easily lead to effective dephasing of resonant molecules and can be considered as a depopulation effect. For the condition on resonance, which is usually the case for saturation spectroscopy, the striction force can be severely enhanced as the dynamic polarizability changes significantly near the molecular resonance.
Recently the effect of standing waves on the vibrational spectrum of HD~\cite{Lv2022} and H$_2$~\cite{Jozwiak2022} was considered.

\begin{figure}[t]
\begin{center}
\includegraphics[width=0.98\linewidth,height=0.2\textheight]{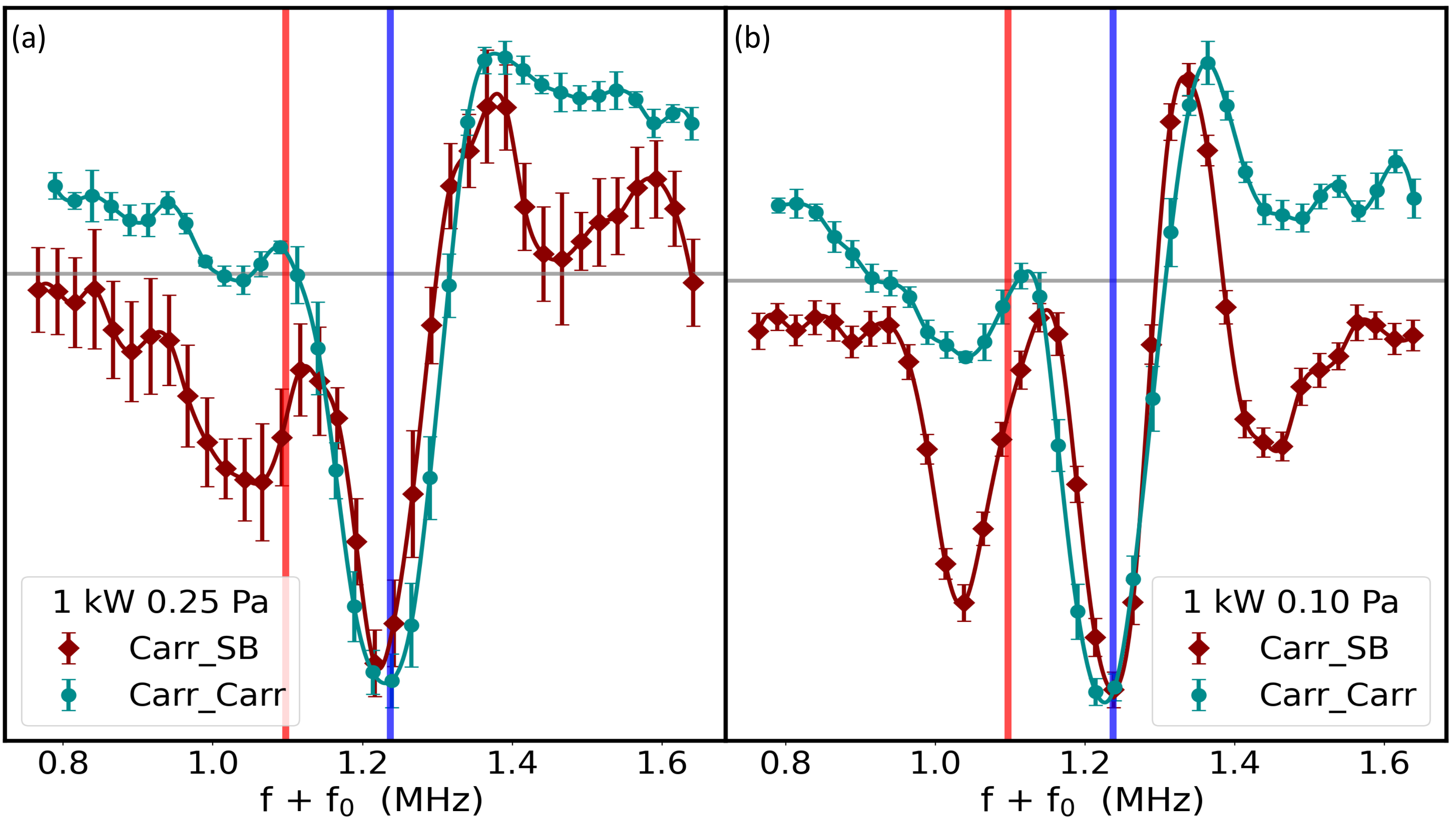}
\caption{\label{CC+CS}
Comparison of NICE-OHMS signals for Carrier-Carrier and Carrier-Sideband (FSR/2 detuned) schemes at 1 kW intracavity power and at (a)  0.25 Pa and (b) 0.10 Pa. 
The sideband (probe) has been kept to a low power of 5 W, equivalent to 0.5\% of the carrier (pump) power. 
Note that the blue line corresponds to the measured frequency of the generic Lamb dip as in Fig.~\ref{H2_S-spectra}.
}
\end{center}
\end{figure}

In order to mitigate the possible effects of the on-resonance strong standing wave and simultaneously prolong the transit time through the intracavity laser beam, Doppler-detuned measurements were performed at FSR/2 detuning, as shown in Fig.~\ref{NO_signals}(b). 
A direct overlap of the two different NICE-OHMS signals (the central carrier-carrier resonance and the FSR/2 detuned carrier-sideband) is accomplished by correcting for the detuning frequency (Fig.~\ref{CC+CS}). 
Comparison between both measurement schemes are made and a blue and red marker for the supposed recoil positions are added, where the blue line indicates the extracted frequency position for the 'generic Lamb dip'. 
This shows that under reduced probe (sideband) powers the red recoil component increases in amplitude, from which can be concluded that the 'generic Lamb dip' at low powers is composed of the blue recoil component only. 
Moreover, at reduced pressure and very low sideband amplitude, the red recoil component is nearly fully recovered. 
Note that at the powers of 1 kW the line shapes of individual components become asymmetric. 
The summation of the unequal intensity recoil components then causes an apparent frequency shift of the weakest (red) component.

From these systematic studies we conclude that the observed generic Lamb dip corresponds to the blue recoil shifted component. 
Correcting for the recoil shift of 70 kHz, and taking into account the contributions to the overall uncertainty (7.3 kHz statistical, with pressure and power effects included, and calibration 1 kHz) the frequency of the S(0) (2-0) quantum transition in H$_2$ is determined at 252\,016\,361\,164\,(8) kHz, with a relative precision of $3 \times 10^{-11}$ representing the most accurate determination of a vibrational splitting in a hydrogen isotopologue~\cite{Fast2020}.

Comparing with the best theoretical result for the S(0) transition frequency, obtained via the approach of non-adiabatic perturbation theory (NAPT)~\cite{Komasa2019} and computed with the H2SPECTRE code~\cite{SPECTRE2022} the experimental result is higher by 2.6 MHz.
The uncertainty from this NAPT approach is 1.6 MHz, hence the deviation between experiment and theory is at $1.6\sigma$, and determined by the $E^{(5)}$ leading order QED-term. 
Part of this $E^{(5)}$ term was recently recomputed~\cite{Silkowski2023} but the issue of systematic discrepancies for vibrational splittings in HD, at the level of $1.9\sigma$, was not resolved.
Now deviations of a similar size are found for the homonuclear H$_2$ species.
For the binding energy of two particular levels in H$_2$, $J=0,1$ in $v=0$, 
separate and dedicated calculations of relativistic and QED corrections were carried out employing nonadiabatic explicitly correlated Gaussian wave functions, yielding an accuracy of 0.78 MHz~\cite{Puchalski2019b}. 
The present experimental results pose a challenging test bench for such advanced theoretical approach.

As an outlook we note that the lifetimes of all rovibrational levels in the H$_2$ electronic ground state exceed $10^5$ s~\cite{Black1976}, thus allowing in principle for metrology of 20-digit precision if the natural lifetime limit can be reached. 
This will push tests of molecular quantum electrodynamics and searches for physics beyond the Standard Model~\cite{Ubachs2016} to the extreme. 
The present experiment signifies a step in that direction.

Financial support from the Netherlands Organisation for Scientific Research (NWO), via the Program “The Mysterious Size of the Proton” is gratefully acknowledged. 
We thank several members of the Quantum Metrology \& Laser Applications group at VU Amsterdam (Edcel Salumbides, Max Beyer, Kjeld Eikema, Jeroen Koelemeij, Yuri van der Werf, Hendrick Bethlem) for helpful discussions.

%

\end{document}